\documentclass[a4paper,showpacs,showkeys,twocolumn,floatfix,prd,nofootinbib]{revtex4}

\usepackage{amsmath}
\usepackage{amssymb}
\usepackage{color}
\newcommand{\be}{\begin{equation}}
\newcommand{\ee}{\end{equation}}

\begin{document}


\title{Baryogenesis via Leptogenesis in an inhomogeneous universe}

\author{A. Kartavtsev}
\email[Email: ]{akartavt@mpi-hd.mpg.de}
\affiliation{Max--Planck Institut f\"ur Kernphysik, Saupfercheckweg 1,
 69117 Heidelberg, Germany}

\author{D. Besak}
\email[Email: ]{dbesak@physik.uni-bielefeld.de}
\affiliation{Fakult\"at f\"ur Physik, Universit\"at Bielefeld, 
33501 Bielefeld, Germany}

\renewcommand{\labelitemi}{--}
\newcommand{\pp}{p}
\newcommand{\px}{{\rm x}}
\newcommand{\nablau}{\nabla\hspace{-0.7mm}u}

\begin{abstract}
We investigate the influence of primordial perturbations of
the energy density and space--time metric on the generation 
of the lepton and baryon asymmetries. In the weak and strong 
washout regimes baryon isocurvature  perturbations with amplitudes 
of the same order as those of the CMB perturbations are generated 
on scales of order of the respective Hubble scale. They are, however,
completely washed out by baryon and photon diffusion at the 
later stages of the universe's evolution.
\end{abstract}

\maketitle

\section{\label{introduction}Introduction.}
Recent oscillation experiments \cite{Ahn:2006zza,Ahmad:2002jz}
have confirmed that the neutrinos have small but nonvanishing 
masses. An observation of neutrinoless double beta decay 
by the future experiments 
\cite{Schonert:2005zn} would hint toward the existence of 
very heavy right--handed Majorana neutrinos 
which generate naturally small neutrino masses via the 
see--saw mechanism. The existence of the heavy neutrinos can 
also explain \cite{Fukugita:1986hr} the observed baryon 
asymmetry of the universe since the three Sakharov conditions
\cite{Sakharov:1967dj} are easily  fulfilled in the 
standard model supplemented by the right--handed
Majorana neutrinos.

The see--saw and baryogenesis via leptogenesis mechanisms
are very attractive from a theoretical point of view. Unfortunately, 
there is still no compelling evidence for the existence of the 
heavy right--handed Majorana neutrinos. Apart from the neutrinoless double beta 
decay experiments one can search for its signatures  in accelerator 
experiments or in lepton flavor violating decays 
\cite{PhysRevLett.45.1908,Ilakovac:1994kj}. However the 
corresponding amplitudes are highly suppressed by the large 
masses of the right--handed neutrinos.

During leptogenesis, processes with right--handed neutrinos 
in initial, intermediate or final states are not suppressed.
In a completely homogeneous and isotropic universe the outcome of 
baryogenesis via leptogenesis is a single number --- the baryon--to--photon 
ratio, which is constant over space. 
In an inhomogeneous universe the primordial perturbations of the 
energy density and space--time metric modulate leptogenesis and
the generated baryon--to--photon ratio depends on the spatial coordinates. 
Thus instead of a single number we get a \textit{function} of space coordinates
which can, in principle, provide us with some additional 
information on the model parameters. 

A general perturbation with arbitrary fractional overdensity in each
component can be represented as a sum of adiabatic 
and isocurvature perturbations. For a purely isocurvature perturbation 
the sum of the fractional overdensities is zero. For a purely adiabatic 
perturbation the fractional overdensity in each matter component (baryons, 
photons, etc.) is the same and hence the baryon--to--photon ratio is constant 
over space. Consequently, a deviation of the baryon--to--photon ratio 
from its average value amounts to a baryon isocurvature perturbation.
We estimate here amplitude and scale dependence of the 
generated perturbations.

Baryon isocurvature perturbations are known to affect primordial 
nucleosynthesis. The abundances of the produced elements sensitively 
depend on the local baryon--to--photon ratio and the final average
abundances can be considerably different from those in the homogeneous
universe \cite{Yang:1983gn,Boesgaard:1985km,PhysRevD.39.2893,KurkiSuonio:1989en}.
They also affect the spectrum of the cosmic microwave background (CMB) radiation. 
Purely isocurvature perturbations produce a series of peaks in the 
CMB spectrum whose angular scales are approximately in the ratio 
$1:3:5\ldots$, whereas purely adiabatic density perturbations produce 
peaks whose angular scales are in the ratio $1:2:3\ldots$ \cite{HuWhite:1996}. 
Current large--scale observations are consistent with the primordial 
perturbations being adiabatic. 
It is however possible that the upcoming experiments will find a small admixture of the 
isocurvature perturbations at scales smaller than 10 Mpc  \cite{Hinshaw:2008kr,Komatsu:2008hk}; in particular the Planck 
satellite will be able to limit the amplitudes of isocurvature 
modes to less than 10\% of the adiabatic mode \cite{Bucher:1999re,Bucher:2000kb}.
The isocurvature component
modifies the sound velocity in the plasma and the initial conditions 
for the CMB \cite{Mukhanov:2005sc}, and is expected to ``smear out'' 
the spectrum, i.e. change height and width of the peaks.

The outline of the paper is as follows. In section \ref{boltzmann}
we derive differential and integral forms of the Boltzmann equation
and prove the gauge invariance of the latter one. The derived integral
Boltzmann equation is used to analyze leptogenesis in section 
\ref{inhomogeneous}, where the generated lepton asymmetry 
is related to the adiabatic energy density and metric perturbations 
and possible velocity perturbation in the heavy neutrino 
component. Finally we summarize our results in section \ref{conclusions}.

\section{\label{boltzmann}The Boltzmann equation}

The general form of the Boltzmann equation is \cite{Dodelson}
\be
\label{generalboltzmann}
\frac{\text{d}f}{\text{d}\lambda} = \hat{C}[f]
\ee
where $\lambda$ is the affine parameter along a geodesic 
and $\hat{C}$ the collision operator acting on the one--particle 
distribution function $f$. 

Since the Boltzmann equation
treats particles as on--shell states, $f$
is a  function of space--time coordinates $x^\alpha$ and three independent 
components of the particle momentum $P_i$, or, alternatively, 
the particle kinetic energy $E$ and unit momentum vector
$\hat{\pp}_i$. Thus we can rewrite the left--hand side 
of \eqref{generalboltzmann} using the chain rule 
\begin{align}
\label{boltzmannchain}
\frac{\partial f}{\partial x^0}+
\frac{\partial f}{\partial x^i} \frac{P^i}{P^0}+
\frac{\partial f}{\partial E}\frac1{ P^0}\frac{dE}{d\lambda}+
\frac{\partial f}{\partial \hat{\pp}_i}\frac1{P^0}\frac{d\hat{\pp}_i}{d\lambda}=
\frac{\hat{C}[f]}{P^0}.
\end{align}
The derivatives of $E$ and $\hat{\pp}_i$ can be calculated using 
the geodesic equation; the result depends on the space--time metric.

The space--time metric of the early universe, which was almost
homogeneous and isotropic, can be split into a background part 
\begin{equation} 
g^0_{\mu\nu}=a^2(\eta){\rm diag}(1,-1,-1,-1),
\end{equation}
where $\eta$ is the conformal time, and a small perturbation which reads \cite{Mukhanov:2005sc}
\begin{equation}
\label{deltagmunu}
\delta g_{\mu \nu } = a^2(\eta)\left( 
\begin{array}{cc}
   2\phi  &  \mathcal{B}_{;i}   \\
    \mathcal{B}_{;i}  & 2\left(\psi \delta_{ij}  + \chi_{ij}  \right)  \\
\end{array}  
\right)
\end{equation}
where $a$ is the cosmic scale factor, $\phi$, $\psi$ and $\mathcal{B}$ are scalar functions,
and $\chi_{ij}$ defined by
\begin{equation}
\chi_{ij}=\left(\partial_i\partial_j-\frac{\delta_{ij}}3 \partial_k\partial_k\right)\mathcal{E}
\end{equation}
is traceless.  Note that the definition of $\psi$ used here differs 
by the $\frac13\partial_k\partial_k \mathcal{E}$ term from that used in
\cite{Mukhanov:2005sc}. This additional term will appear in the equation 
for the gauge transformation \eqref{psitransform} of $\psi$ later on. 

A nonzero shift function $\mathcal{B}$ means that comoving worldlines 
and worldlines orthogonal to hypersurfaces of constant time are 
not collinear and we are dealing with a locally nonorthogonal 
coordinate system \cite{LiddleLyth}. We avoid the associated 
unnecessary complications by setting $\mathcal{B}=0$  which is fulfilled, 
for instance, in all synchronous gauges and the longitudinal gauge.

Isotropy of the background space--time implies that the zero--order 
phase--space distribution functions are independent of the direction of 
the momentum vector, so that $\frac{\partial f}{\partial \hat{\pp}_i}$ 
vanishes in the homogeneous universe defined by $g^0_{\mu\nu}$ and is of first 
order in the inhomogeneous universe defined by \eqref{deltagmunu}. Isotropy also 
implies that in the homogeneous universe the unit momentum vector 
is conserved, so that in the inhomogeneous universe 
$\frac{d\hat{\pp}_i}{d\lambda}$ is of first order as well. Consequently, the
product of the two is of second order and the last term on 
the left--hand side of \eqref{boltzmannchain} can be neglected. 
Then the Boltzmann  equation takes the form
\begin{align}
\label{boltzmannfinal}
\frac{\partial f}{\partial x^0} &+\frac{\partial f}{\partial x^i}
\frac{P^i}{P^0}-
\frac{\partial f}{\partial E}
\left(\frac{\pp^2}{E}
\left[
H-\dot{\psi}-\dot{\chi}_{ii}\,\hat{\pp}_i^2
\right]+\phi_{,i}\, \pp \hat{\pp}_i
\right) \hfill \nonumber \\
&=\sqrt{g_{00}}\,\frac{\hat{C}[f]}{E}\hfill \nonumber \\ 
\sqrt{g_{00}} &=a(1+\phi). \hfill
\end{align}

The expression in the round brackets is (up to a sign) 
$\frac1{ P^0}\frac{dE}{d\lambda}$ calculated to linear order 
in the small perturbations of the metric, and $H\equiv \frac{\dot a}{a}$
is the Hubble parameter. In the longitudinal gauge $\mathcal{B}=\chi_{ij}=0$
and the Boltzmann equation \eqref{boltzmannfinal}  reverts to that 
derived in \cite{Dodelson}.

As we are only interested in the total generated  asymmetry,
we integrate the Boltzmann equation \eqref{boltzmannfinal} over 
the phase space. Integration of the first term on the left--hand side 
gives the derivative of the particle number density $n$ with respect to the 
time coordinate $x^0$.

Since for a massive particle $\frac{P^i}{P^0}=\frac{U^i}{U^0}=V^i$,  
the second term on the left--hand side can be rewritten in the form 
\begin{equation}
\label{secondterm}
\int \frac{\partial f}{\partial x^i}\frac{P^i}{P^0}
d\Omega_\pp\,=
\frac{\partial }{\partial x^i}
\int f V^i d\Omega_\pp\,=
\frac{\partial (n v^i) }{\partial x^i}
\end{equation}
where $v^i$ denotes the macroscopic three--velocity of the gas.
In the early universe the macroscopic three--velocity is a 
first--order quantity and therefore  (up to second order 
corrections) is proportional to the corresponding macroscopic 
four--velocity: $v^i\approx a u^i$ (recall  that $u^0\approx a^{-1}$).

Since $\chi_{ij}$ is of first order, one can integrate
the fifth term in \eqref{boltzmannfinal} over the isotropic 
zero--order phase--space distribution function. In this approximation 
the integration   gives the same result for any 
$i$ and therefore (recall that $\chi_{ij}$ is traceless) 
this term vanishes to first order. 

Analogously, since $\phi$ is of first order, one can again integrate 
the last term on the left--hand side of \eqref{boltzmannfinal}
over the phase--space using the zero--order phase--space distribution 
function; the integral vanishes due to the isotropy of the 
distribution function.

Upon integration by parts, the remaining terms take the form
\begin{equation}
\label{thirdterm}
3(H-\dot{\psi})n=\frac{n}{\sqrt{-g_3}}\frac{\partial \sqrt{-g_3}}{\partial x^0},
\quad 
\sqrt{-g_3}=a^3(1-3\psi).
\end{equation}
Using the fact that $\chi_{ij}$ is traceless  it is 
straightforward to check that $g_3$ is the determinant 
of the spatial part of the metric.

Collecting all the terms  we obtain the integral form of the Boltzmann 
equation
\begin{equation}
\label{boltzmannintegr}
\frac1{\sqrt{-g_3}}\frac{\partial (n\sqrt{-g_3})}{\partial x^0}+
a\frac{\partial (n u^i) }{\partial x^i}=
\sqrt{g_{00}}\int \frac{\hat{C}[f]}{E} d\Omega_\pp.
\end{equation}

Equation \eqref{boltzmannintegr} is a partial differential equation 
which is rather difficult to solve. However, if we neglect some further
second--order
terms, it can be substantially simplified. Since $u^i$ as well 
as the gradient of $n$ are of first order, their product can 
be neglected, so that $(n u^i)_{,i}\simeq n
u^i_{,i}$. Next we introduce 
\begin{equation}
\label{newY}
\Upsilon\equiv n\sigma\,,\quad \sigma\equiv \sqrt{-g_3}\, 
\exp{\left({\textstyle \int} u^i_{,i}\, adx^0\right)}.
\end{equation}
As can be checked by direct substitution, equation \eqref{boltzmannintegr}
is equivalent (up to the second order terms) to the following equation for 
$\Upsilon$
\begin{align}
\label{eqforupsilon}
\frac1{\sqrt{g_{00}}}\frac{\partial \Upsilon}{\partial x^0}= 
\sigma \int \frac{\hat{C}[f]}{E} d\Omega_\pp\,
\end{align}
where all the terms are evaluated at the \emph{same} point in space.

Equation \eqref{eqforupsilon} is an \emph{ordinary} differential equation
with respect to the time coordinate.  Any dependence of the generated lepton
asymmetry on the space coordinates is ``hidden'' in the
dependence of the metric and macroscopic parameters  on the space coordinates.
This substantial simplification is possible because we 
limit our analysis to linear perturbations here. 

From \eqref{eqforupsilon} it is clear that $\Upsilon$ is conserved if 
the collision terms vanish;  that is, in the inhomogeneous universe 
$\Upsilon$ replaces the particle number density per comoving volume 
$Y$. 

Let us also note  that equations \eqref{boltzmannintegr} and  \eqref{eqforupsilon} 
describe the development of the particle number density of a particular species and 
can be used in a universe with an arbitrary mixture of  adiabatic and 
isocurvature perturbations. 

The solutions of the Einstein equations give macroscopic parameters
like the temperature as functions of the time coordinate. Since 
the collision terms depend explicitly on the
inverse temperature $z=\frac{M_1}{T}$ (where $M_1$ is mass 
of the lightest Majorana neutrino) \cite{PhysRevD.58.113009} 
rather than on the time coordinate $x^0$, it is convenient to rewrite 
\eqref{eqforupsilon} as follows:
\begin{equation}
\label{boltzmannzprime}
\Upsilon'=\left(\frac{\sqrt{g_{00}}}{\dot{z}}\right)\sigma
\int \frac{\hat{C}[f]}{E} d\Omega_\pp\,.
\end{equation}
The dot denotes differentiation with respect to the time coordinate,
whereas the prime denotes differentiation with respect to $z$.

The analysis of processes in the early universe is often 
complicated by the gauge choice. It is therefore important 
to check if the derived integral form of the Boltzmann 
equation is invariant with respect to the gauge transformations 
which preserve $\mathcal{B}=0$. Let us consider a transformation
from synchronous to longitudinal coordinates. Small perturbations 
of the metric and  fluid velocity in the longitudinal gauge are related to 
those in the synchronous gauge by \cite{Mukhanov:2005sc}
\begin{subequations}
\label{phipsi}
\begin{align}
\label{psitransform}
\psi_{(l)}&=\psi_{(s)}-\frac{\dot{a}}{a}\dot{\mathcal{E}}_{(s)}-
\frac13\partial_k\partial_k \mathcal{E}_{(s)},\\
\label{utransform}
u_{i(l)}&=u_{i(s)}+a\,\partial_i \dot{\mathcal{E}}_{(s)}.
\end{align}
\end{subequations}
The same value of the proper time $t$ (which is gauge invariant by 
definition) corresponds in the two gauges 
to two different values of the time coordinate: $x^0_{(s)}$ in 
the synchronous gauge and $x^0_{(l)}$ in the longitudinal
one. Using the transformations laws for $\cal B$ and $\cal E$
\begin{equation}
{\mathcal B}_{(l)}={\mathcal B}_{(s)}+\dot{\zeta}+x^0_{(s)}-x^0_{(l)},\quad
{\mathcal E}_{(l)}={\mathcal E}_{(s)}+\zeta
\end{equation}
(see \cite{Mukhanov:2005sc} for more details) and taking into account 
that $\cal B$ is zero in both gauges and $\cal E$ is zero in the 
longitudinal gauge, we obtain 
\begin{equation}
\label{x0transform}
\Delta x^0=x^0_{(s)}-x^0_{(l)}=\dot{\mathcal{E}}_{(s)}.
\end{equation}
Using equations \eqref{thirdterm}, \eqref{psitransform} and \eqref{x0transform} 
we find the following transformation law for the determinant of the spatial 
part of the metric
\begin{equation}
\label{g3transormation}
\sqrt{-g_3}_{(l)}(x^0_{(l)})=\sqrt{-g_3}_{(s)}(x^0_{(s)})
\left(1+\partial_k\partial_k \mathcal{E}_{(s)}\right).
\end{equation}
Taking the divergence of \eqref{utransform} and multiplying by the 
scale factor we obtain 
\begin{equation}
\label{dudxtransformation}
a u^i_{,i(l)}(x^0_{(l)})= a u^i_{,i(s)}(x^0_{(s)})-
\partial_k\partial_k\dot{\mathcal{E}}_{(s)}.
\end{equation}
After   integration over the time coordinate in \eqref{newY} and 
asymptotic expansion of the exponent,
the second term on the right--hand side of \eqref{dudxtransformation} 
cancels the analogous term in \eqref{g3transormation}, 
so that $\sigma_{(l)}(x^0_{(l)})=\sigma_{(s)}(x^0_{(s)})$. 
In other words, $\sigma$ is invariant under the gauge transformation.
Three--dimensional scalars like the particle number density are 
gauge invariant as well: $n_{(l)}(x^0_{(l)})=n_{(s)}(x^0_{(s)})$  \cite{Mukhanov:1990me}. Together with the gauge invariance of $\sigma$ this implies the 
gauge invariance of $\Upsilon$.

Another important gauge invariant three--dimensional scalar is the 
energy density $\rho$
\begin{equation}
\rho=T^0_0={\textstyle \int} f \sqrt{-g}\,P^0  dP^1dP^2dP^3=
{\textstyle \int} fEd\Omega_p.
\end{equation}
As can be shown by a direct calculation, for a gas of massless particles
distributed according to the Boltzmann distribution
the leading contribution to the energy density due to a
small    macroscopic velocity $\vec u$ is proportional 
to $\vec{u}\,^2$, i.e.   is of second order and can be neglected.
Thus $\rho= const.\cdot T^4$, and the temperature $T$ (as well as the
inverse temperature $z$) is gauge--invariant. 

Therefore the 
left--hand side of \eqref{boltzmannzprime} is gauge invariant since
the derivative of one invariant quantity with respect to another invariant
quantity is also invariant.

On the right--hand side of \eqref{boltzmannzprime} the integral 
of the collision terms over the phase space is also invariant under 
the gauge transformation \footnote{Perturbations of the space time 
metric may in principle lead to a change of decay widths and cross 
sections of the scattering processes. These effects, however,
are  not considered here. Following  common practice we assume 
that the widths and cross sections are given by the same expressions as 
those in  vacuum. Then the perturbations only affect the thermally averaged 
decay widths and scattering cross sections through the modification of the 
temperature and the related modification of the equilibrium phase--space
distribution functions.}. Finally, since 
\begin{equation}
\label{zdot}
\frac{\sqrt{g_{00}}}{\dot{z}}\equiv 
\frac{\sqrt{g_{00}}}{\left(\frac{\partial z}{\partial x^0}\right)}=
\left(\frac{\partial z}{\partial t}\right)^{-1}
\end{equation}
and the (inverse) temperature $z$ as well as the proper time $t$ are gauge 
invariant, the ratio \eqref{zdot} is also invariant. This completes our 
proof of the  invariance of the integral  Boltzmann equation under 
the gauge transformation \eqref{phipsi}. 

Instead of the transformation \eqref{phipsi} between synchronous and 
longitudinal gauge, we could have considered a transformation between 
two different gauges which fulfill $\mathcal{B} = 0$. The calculation 
would differ in only $\cal E$ being now different from zero in both 
gauges and appearing also on the left--hand side of equations 
\eqref{g3transormation} and \eqref{dudxtransformation}, which however
does not change the conclusion concerning the invariance of $\sigma$. 
Therefore, we may
conclude that equation \eqref{boltzmannzprime} is invariant not only under the
transformation between longitudinal and synchronous gauge, but under all
gauge transformations that preserve the gauge condition $\mathcal{B}=0$.

\section{\label{inhomogeneous}Inhomogeneous leptogenesis}

To calculate the lepton asymmetry generated in the decays of the 
heavy Majorana neutrinos one has to solve a coupled system of  
Boltzmann equations for the heavy neutrino and lepton 
number densities. 

To be able to perform this calculation analytically we neglect the  
contribution of the heavy neutrinos to the energy density of the universe. 
This is justified by the smallness of the associated number of effective 
massless degrees of freedom as compared to that of all the SM species.
The two--body scattering processes and the flavor effects are also neglected
for simplicity. In the simplest scenario of inflation isocurvature perturbations
in the gas of the SM species are suppressed and we also neglect them.
In particular, this implies that $\sigma$ is the same 
for leptons and photons so that $\eta_L\equiv n_L/n_\gamma=\Upsilon_L/\Upsilon_\gamma$.
The possible deviation of the macroscopic velocity of the heavy neutrino gas 
from that of leptons (i.e.  velocity perturbation in the 
right--handed neutrino component) is parameterized by
$r_u\equiv \frac{\sigma_\ell}{\sigma_\Psi}-1\approx \int (u^i_\ell-u^i_\Psi)_{,i}\,a dx^0$.

\subsection{Approximate analytical solution of the system of Boltzmann equations}

The collision terms are determined by the coupling of the 
Majorana  neutrino to leptons and the Higgs
\begin{equation}
\mathcal{L}=N(\bar{\ell}\hat{\lambda}^\dagger\tilde{h})-
\textstyle{\frac12}\bar{N}\hat{M}N^c+{\rm h.c.}
\end{equation}
where $\ell$ denotes the leptons, $\tilde{h}=i\sigma_2 h^\dagger$
is the charge conjugate Higgs doublet, and $N$ are the components of 
physical Majorana neutrino field $\Psi=\frac1{\sqrt{2}}(N+N^c)$. 

Under the usual assumption that the phase--space distribution function
of the Majorana neutrino is proportional to the equilibrium one, the
contribution of the decay ($\Psi\rightarrow \ell h^\dagger$) 
and the inverse decay ($\ell h^\dagger \rightarrow \Psi$)  processes
to the  lepton asymmetry reads \cite{PhysRevD.58.113009}
\begin{equation}
\int \frac{\hat{C}[f_L]}{E} d\Omega_\pp\,=
\langle \Gamma_{\Psi_1} \rangle\left[
\varepsilon(n_\Psi-n^{\text{eq}}_\Psi)-\frac{n^{\text{eq}}_\Psi}{n^{\text{eq}}_\ell}n_\ell
\right]
\end{equation}
where $\varepsilon$ is the usual \textit{CP} asymmetry parameter and
 $\langle \Gamma_{\Psi_1} \rangle$ is the thermally averaged 
heavy neutrino decay width. Integrating over the equilibrium 
distribution function and neglecting a second--order contribution
due to the small macroscopic velocity \footnote{In the gas rest 
frame the thermally averaged decay width depends only on temperature.
After a Lorentz boost to the initial reference frame the  
decay width acquires the Lorentz factor, which differs from unity 
only by a second order term.} we find  
\begin{equation}
\langle \Gamma_{\Psi_1} \rangle=\Gamma_{\Psi_1}\frac{K_1(z)}{K_2(z)},\quad
\Gamma_{\Psi_1}=\frac{(\lambda\lambda^\dagger)_{11}}{8\pi}M_1\,,
\end{equation}
where $K_1$ and $K_2$ are the modified Bessel functions. The equilibrium
number densities are obtained by integration of the corresponding
Maxwell--Boltzmann distribution functions over the phase space 
\begin{equation}
\label{eqnumdens}
n^{\text{eq}}_\Psi=\frac{T^3}{\pi^2}z^2K_2(z),\quad 
n^{\text{eq}}_\ell=4N\frac{T^3}{\pi^2}
\end{equation}
where $N=3$ is the number of generations. One should keep in mind 
that $T$ refers to leptons and the Higgs; the right--handed neutrino 
is out of equilibrium and does not have a definite temperature. 

The same processes also contribute to the right--hand side of the 
Boltzmann equation for the Majorana number density \cite{PhysRevD.58.113009}
\begin{equation}
\int \frac{\hat{C}[f_\Psi]}{E} d\Omega_\pp\,=-
\langle \Gamma_{\Psi_1} \rangle \left(n_\Psi-n^{\text{eq}}_\Psi\right).
\end{equation}

In the homogeneous and isotropic universe dominated by radiation
(see \cite{Mukhanov:2005sc} for more details)
\begin{equation}
\sigma=\sqrt{-g_3}=a^3 \propto T_0^{-3},\quad 
\left(\frac{\sqrt{g_{00}}}{\dot{z}}\right)=
\frac{z_0}{\cal H}
\end{equation}
where ($z_0$) $T_0$ is the (inverse) background temperature 
and ${\cal H}\equiv H(M_1)$ is value of the Hubble parameter at a temperature 
equal to the mass of the lightest Majorana neutrino. In the 
inhomogeneous universe we tentatively write 
\begin{equation}
\label{sigmaansatz}
\sigma\propto T^{-3} (1+r_\sigma),\quad 
\left(\frac{\sqrt{g_{00}}}{\dot{z}}\right)=
\frac{z}{\cal H}(1+r_{\dot{z}})
\end{equation}
where $r_\sigma$ and $r_{\dot{z}}$ are first--order gauge--invariant\footnote{The gauge invariance of $r_\sigma$ and $r_{\dot z}$
follows from the gauge invariance of the (inverse) temperature
and of $\sigma$ and $\frac{\partial z}{\partial t}$.} functions. 
In what follows we will use $r_{\sigma_\Psi}$ and $r_{\sigma}$
for the heavy neutrino and SM species, respectively.
Introducing  finally 
$\kappa\equiv \Gamma_{\Psi_1}/{\cal H}$, $\Delta\equiv
\Upsilon_\Psi-\Upsilon^{\text{eq}}_\Psi$ and 
\begin{equation}
\gamma_D(z)\equiv \frac{K_1(z)}{K_2(z)},\quad
\gamma_L(z)\equiv \frac{z^2}{4N}K_2(z)
\end{equation}
we obtain the following 
system of  Boltzmann equations for the lepton asymmetry 
and Majorana neutrino number density 
\begin{subequations}
\label{boltzmanneqsanalyt}
\begin{align}
\label{boltzmanneqsanalyt1}
\Delta'&=-\kappa z\,(1+r_{\dot{z}})\,\gamma_D\Delta-\Upsilon^{'\text{eq}}_\Psi\\
\label{boltzmanneqsanalyt2}
\Upsilon'_L&=\kappa z\,(1+r_{\dot{z}})\left[\varepsilon(1+r_u)\gamma_D\Delta
-\gamma_L\Upsilon_L\right] 
\end{align}
\end{subequations}
where the relation  $r_u=r_{\sigma}-r_{\sigma_\Psi}$ has been 
used. If $r_{\sigma}$, $r_{\sigma_\Psi}$  and $r_{\dot{z}}$ vanish, 
the resulting system coincides with that in the homogeneous 
universe and the generated asymmetry depends only on  
$\varepsilon$ and $\kappa$.

In the weak washout regime ($\kappa\ll 1$) the deviation from 
equilibrium is large and   washout processes play almost no role,
so that $\Upsilon_L'\approx-\varepsilon(1+r_u)\Upsilon_\Psi'$. Together 
with the initial conditions $\Delta(0)=\Upsilon_L(0)=0$ this gives 
upon integration by parts and use of the zero--order solution of 
\eqref{boltzmanneqsanalyt1}, 
$\Upsilon^0_\Psi(z)=
\Upsilon_\Psi(0)\exp(-\kappa z^2/2)$, (see \cite{Kolb:1990vq})
\begin{equation}
\label{Yweak}
\Upsilon_L\approx\Upsilon^0_L \left[1+r_\sigma(0)+
 \int r'_u \exp{(-\kappa z^2/2)} dz\right],
\end{equation}
where the unperturbed value of the generated asymmetry in the 
weak washout regime is given by 
\begin{equation}
\Upsilon^0_L=\frac{2\varepsilon}{\pi^2}.
\end{equation}
 
In the strong washout regime ($\kappa\gg 1$)  the deviation from 
equilibrium is small, so that the final lepton asymmetry is independent 
of the initial conditions and determined by the dynamics of leptogenesis. 
In this regime the density of Majorana neutrinos closely tracks its 
equilibrium value, which implies  $\Delta'\simeq 0$ \cite{Kolb:1990vq}. 
Substitution of \eqref{boltzmanneqsanalyt1} into \eqref{boltzmanneqsanalyt2} 
then gives 
\begin{equation}
\Upsilon'_L=-\varepsilon(1+r_u)\Upsilon^{'\text{eq}}_\psi-
\kappa z\,(1+r_{\dot z})\gamma_L \Upsilon_L.
\end{equation}
The formal solution of this equation satisfying the initial conditions 
$\Delta(0)=\Upsilon_L(0)=0$ reads 
\begin{equation}
\label{formalsolution}
\Upsilon_L=-\varepsilon \int_0^z (1+r_u) 
\Upsilon^{'\text{eq}}_\Psi \exp\left(
-{\textstyle \int}^z_{z'}\kappa z'' (1+r_{\dot z})\gamma_Ldz''
\right)dz'.
\end{equation}
For $z\gg 1$ one can use the large--argument asymptotics of the 
Bessel functions $K_1$ and $K_2$, so that approximately  
\begin{align}
\label{upsilonprime}
-\Upsilon^{'\text{eq}}_\psi(z) & \approx [1+r_{\sigma_\Psi}(z)-r'_{\sigma_\Psi}(z)]\sqrt{\frac{z^3}{2\pi^3}}
\exp(-z).
\end{align}
The arguments of the exponents in \eqref{formalsolution} and 
\eqref{upsilonprime} combine to a function which has a sharp 
peak at $z=z_f$, where $z_f$ is the solution of
\begin{equation}
\label{zfequation}
\kappa z_f\gamma_L(z_f)[1+r_{\dot z}(z_f)]=1.
\end{equation}
Consequently one can apply the method of 
steepest descent to estimate the asymptotic value of this integral.
Approximating $\gamma_L$ by its large--argument asymptotics and 
keeping only the terms of highest order in $z_f$ we obtain
\begin{subequations}
\label{estimates}
\begin{align}
-(\kappa z\gamma_L[1+r_{\dot z}])'_{z=z_f}&\approx 1-r'_{\dot z}(z_f),\\
\int_{z_f}^\infty \kappa z\gamma_L [1+r_{\dot z}]\, dz&\approx 1+r'_{\dot z}(z_f).
\end{align}
\end{subequations}
A deviation of $r_{\dot z}$ from zero leads to a deviation  of $z_f$
from its unperturbed value. Using \eqref{zfequation} one can show 
that the shift is given by $z_f-z_{f_0}\approx r_{\dot z}(z_{f})$,
where $z_{f_0}$ denotes value of the freeze--out temperature in the  
$r_{\dot z}=0$ case. 

Using equations \eqref{zfequation} and \eqref{estimates} we get for the
asymptotic value of the asymmetry \footnote{On very 
small scales the thermodynamic quantities rapidly oscillate and the 
solution obtained by the method of steepest descent under the assumption
that the functions under the integral are smooth and do not oscillate 
is not valid.}
\begin{align}
\label{Ystrong}
\Upsilon_L \approx \Upsilon^0_L
\frac{1+r_\sigma(z_f)-r'_\sigma(z_f)+r'_u(z_f)}
{1+(1+z^{-1}_{f})r_{\dot z}(z_f)+\frac12 r'_{\dot z}(z_f)}
\end{align}
where the unperturbed value of the generated asymmetry 
in the strong washout regime is given by 
\begin{equation}
\Upsilon^0_L=\frac{2\varepsilon}{\pi^2}
\frac{N}{\kappa z_{f_0}}\frac{2\sqrt{2\pi}}{\exp(1)}.
\end{equation}
Despite its simplicity, for $\kappa\gtrsim 10$ the analytical 
solution  \eqref{Ystrong} deviates from  the exact numerical 
solution of \eqref{boltzmanneqsanalyt} by only a few percent 
and correctly describes the  dependence of the 
asymmetry on $r_\sigma$ and $r_{\dot{z}}$. 
For large $\kappa$ the inverse freeze--out temperature $z_{f}\gg 1$
and the term proportional to $z_f{\scriptstyle ^{-1}}$  in the denominator can be 
neglected.  

\subsection{The baryon--to--photon ratio}

To complete the analysis we have to evaluate 
$r_\sigma$ and $r_{\dot z}$. Introducing the
dimensionless temperature perturbation $\Theta\equiv \frac{\delta T}{T}$
one can relate $z$ to its background value $z=z_0(1+\Theta)^{-1}$.
Using the known dependence of $z_0$ on the time--coordinate in the 
radiation--dominated universe, $z_0={\cal H}\, x^0 \propto a$, 
and the resulting  relation $\dot{z}_0z^{-1}_0=H$ we obtain for time 
derivative of $z$
\begin{equation}
\frac{dz}{dt}=\frac{1}{\sqrt{g_{00}}}\frac{\partial z}{\partial x^0}
=\frac{\cal H}{z_0(1+\Theta)}\frac{1}{\sqrt{g_{00}}}\left(1-\frac1{H}
\frac{\partial\Theta}{\partial x^0}\right),
\end{equation}
which implies 
\begin{equation}
\label{rzdotexpr}
r_{\dot z}=\phi+2\Theta+H^{-1}\dot{\Theta}\,.
\end{equation}
Expanding the exponent in \eqref{newY} and 
neglecting second--order terms one can write $\sigma$ in
the form
\begin{equation}
\sigma=a^3(x^0)\left(1-3\psi+{\textstyle \int} a u^i_{,i}\,dx^0\right),
\end{equation}
where the scale factor $a$ is related to the  background 
temperature by $a^3\propto T_0^{-3}=T^{-3}(1+3\Theta)$. Consequently
\begin{equation}
r_\sigma=3\Theta-3\psi+{\textstyle \int} a u^i_{,i} dx^0.
\end{equation}

Since $r_\sigma$ and $r_{\dot z}$ are gauge invariant, it is 
sufficient to evaluate them in one particular gauge. In the 
longitudinal gauge solution of the Einstein equations for the Fourier 
components of $\phi$ and $\Theta$  in the radiation--dominated 
universe reads \cite{Mukhanov:2005sc}
\begin{subequations}
\begin{align}
\phi_k&=
C_1\left(\frac{\sin x}{x^3}-\frac{\cos x}{x^2}\right)+
C_2\left(\frac{\cos x}{x^3}+\frac{\sin x}{x^2}\right)\\
\Theta_k&=\frac{C_1}{2}\left[
\left(\frac{2-x^2}{x^2}\right)\left(\frac{\sin x}{x}-\cos x\right)-
\frac{\sin x}{x}
\right]\nonumber\\
&+\,\,C_2\left[
\left(\frac{1-x^2}{x^2}\right)\left(\vphantom{\frac{\sin x}{x}}\frac{\cos x}{x}+\sin x\right)+
\frac{\sin x}{2}
\right]
\end{align}
\end{subequations}
where $x\equiv \frac{kx^0}{\sqrt{3}}$. The initial conditions are 
specified by  $C_1$ and $C_2$, which are functions of $k$.
The divergence of the macroscopic fluid velocity $\upsilon\equiv u^i_{,i}$ 
can be calculated using the Einstein equation for the off--diagonal 
components of the Ricci tensor. In the  absence of anisotropic stress,
$\psi=\phi$, and these simplify to 
$\dot{(a\phi)}_{,i}=4\pi G a^2(\varepsilon_0+p_0) u_i$ \cite{Mukhanov:2005sc}.
After use of the Friedmann equation, $3 H^2=8\pi Ga^2 \varepsilon_0$,
and some algebra we obtain for the corresponding Fourier 
components in the flat radiation--dominated universe 
\begin{equation}
\label{upsilonk}
\upsilon_k=\frac{k^2}{2\dot{a}^2}\dot{(a\phi_k)}.
\end{equation}
Trading the integration with respect to the time coordinate 
for the integration with respect to $x$ and taking into account 
the $\psi=\phi$ equality we find for the Fourier transform of $r_\sigma$\footnote{Equation \eqref{rsigmak} implies that for the photons
$\Upsilon_\gamma=\Upsilon^0_\gamma=\frac{2}{\pi^2}$ is constant in 
the early universe, just as expected. Since photons
are in equilibrium in the early universe, the right--hand side of the 
Boltzmann equation for photons vanishes and $\Upsilon_\gamma$ is conserved.}
\begin{equation}
\label{rsigmak}
r_{\sigma_k}=3\Theta_k-3\phi_k+\int\frac{3x}{2} \frac{\partial }{\partial x}(x\phi_k) dx=0.
\end{equation} 
Using the relation $\dot{\Theta}=\dot{x}\frac{\partial }
{\partial x}\Theta=Hx \frac{\partial \Theta}{\partial x}$ we obtain 
a very simple expression  for the Fourier transform of $r_{\dot z}$
\footnote{A calculation using the solution of the Einstein equations 
for small perturbations in the synchronous gauge given in \cite{Bednarz:1984dn} 
leads to the same expressions for $r_\sigma$ and $r_{\dot z}$; this 
confirms the gauge invariance of the introduced functions.}
\begin{equation}
\label{rzdot}
r_{{\dot z}_k}=\phi_k+2\Theta_k+x \frac{\partial \Theta_k}{\partial x}=
-\frac{x}{2}\left( C_1\sin x +C_2 \cos x\right).
\end{equation}

The lepton to photon number ratio, $\eta_L=\Upsilon_L/\Upsilon_\gamma$, can now 
be easily read off from  \eqref{Yweak} and \eqref{Ystrong}.
The quantity of interest is the departure of the asymmetry 
from its background value, $r_\eta\equiv \eta_L/\eta^0_L-1$. 
Since the baryon asymmetry is linearly related to the lepton 
one, $r_\eta$ also characterizes the baryon isocurvature
perturbations. Neglecting the term proportional to $z^{-1}_f$  in the denominator 
of \eqref{Ystrong}, we find in the strong washout regime \footnote{Since the 
contribution of the heavy neutrinos to the energy density is 
neglected, the baryon--to--photon ratio is related to the baryon--to--entropy 
ratio by a constant factor at this stage of the 
universe's evolution. Later, as the SM species decouple
as they become massive,
the baryon--to--entropy ratio remains constant, wheres the baryon--to--photon 
ratio changes. This change is absorbed into the definition 
of the background value of the asymmetry and does not modify the 
expressions for $r_\eta$.}
\begin{equation}
\label{retastrong1}
r_\eta=r'_u(z_f)-r_{\dot z}(z_f)-{\textstyle \frac12} r'_{\dot z}(z_f).
\end{equation}
The calculation of the CMB anisotropies is performed in  momentum 
space, so that one is interested in the Fourier components of \eqref{retastrong1}. 
The Fourier transform must be evaluated at 
the inverse freeze--out temperature $z_f$; it corresponds to $x_f=z_f {\cal X}$, 
where ${\cal X}\equiv \frac{k {\cal R}}{\sqrt{3}}$ is the value 
$x$ takes at $z=1$, and $\mathcal{R}\equiv \mathcal{H}^{-1}\propto M_{Pl}M_1^{-2}$ 
is the Hubble scale at this stage. The derivative of \eqref{rzdot} with respect 
to the inverse temperature can be obtained upon use of the zero--order 
relation $\frac{\partial}{\partial z}={\cal X}\frac{\partial}{\partial x}$. 
Using the introduced notation we find for the Fourier transform of $r_\eta$ 
in the strong washout regime
\begin{align}
\label{retastrong}
r_{\eta_k}={ \cal X}&\left[{\textstyle \frac{\partial}{\partial x}}r_u+{\textstyle\frac14}
(2C_1\right. z_f-C_2)\sin {\cal X}z_f\nonumber\\
+&\left.{\textstyle\frac14}(2C_2 z_f+C_1)\cos {\cal X}z_f\right].
\end{align}
In this regime the heavy neutrinos are tightly 
coupled to the SM species and it is natural to expect 
that  $r_u$ vanishes. The typical size of the 
perturbations is given by the Hubble radius $\mathcal{R}$.
On larger scales $r_{\eta_k}$ is suppressed by the overall factor 
$\cal X$, which becomes small. On superhorizon scales 
$r_{\eta_k}\approx \frac14\mathcal{X}(2C_2 z_f+C_1)$ and
vanishes as $\kappa\rightarrow \infty$. 

From \eqref{Yweak} it follows that in the weak washout 
regime 
\begin{equation}
\label{retaweak}
r_\eta= \int \left({\textstyle \frac{\partial}{\partial x}}
r_u\right) \exp{(-\kappa x^2/2{\cal X}^2)} dx\,,
\end{equation}
where we have used the relation $x=z{\cal X}$. 
That is, a nonzero $r_\eta$ is generated in this regime only if the
velocity of the right--handed neutrino  deviates from that 
of the SM species. The typical scale and amplitude of  $r_u$ are 
model--dependent. Nevertheless one can
make a plausible guess about these quantities: if perturbations
in the SM and heavy neutrino components are created by the same
mechanism, then the typical amplitude and scale of $r_u$ cannot 
exceed those of the velocity perturbation of the SM species. 
Since we are interested in large--scale perturbations, which 
correspond to small $\cal X$, it is sufficient to know the behaviour 
of $r_u$ in the vicinity of $x=0$ 
\begin{align}
r_u=\sum a_n x^n, \quad a_n\sim {\cal O}(C_{1,2})\,.
\end{align}
If, for example, $u_\psi$ vanishes in the longitudinal 
coordinates\footnote{The analysis  
performed in \cite{Green:2005fa} in longitudinal gauge 
demonstrates that velocity perturbation of the heavy dark 
matter particles are small compared to those of the massless 
species.}, then as follows from \eqref{upsilonk} $a_{-1}=3C_2$, $a_0=\frac12C_1$, $a_2=\frac14C_1 \ldots$\,,
whereas if $u_\psi$  vanishes in synchronous coordinates, then 
$a_1=\frac32 C_2$, $a_4=\frac{1}{32}C_1 \ldots$ (see Eq. (15) in \cite{Bednarz:1984dn}). To integrate 
the terms with negative $n$ we cut the integral at some $x_{min}$ such
that $a_n x_{min}^n\ll 1$ to insure the applicability of the 
perturbative treatment. We furthermore consider only very 
large scales ${\cal X}\ll x_{min}$. Then 
\begin{align}
\label{retaweak2}
r_{\eta_k}= a_1\sqrt{\pi/2\kappa}\,{\cal X}+{\cal O}({\cal X}^2)\,.
\end{align}
That is, in the weak washout regime $r_u$ is also suppressed on large scales.

\subsection{BBN and CMB}

In the radiation--dominated universe the horizon size 
grows much faster than the scale factor, and consequently 
the generated perturbations are not protected from smoothing 
by baryon diffusion at later stages of the universe's evolution.

Let us first consider the strong washout regime. Relation 
\eqref{retastrong} holds for arbitrary mass of the Majorana 
neutrino. For the right--handed neutrino mass $M_1\sim 10^9$ GeV, 
scales of the order of  $\cal R$ at the time of 
leptogenesis correspond to (roughly speaking) $10^{-4}$ m at  
neutrino decoupling. This is much smaller than the baryon diffusion 
length $\sim 200$ m \cite{Schwarz:2003du} at this stage.  
Therefore in this case the perturbations of the baryon to photon ratio
are completely smoothed out before nucleosynthesis starts. 
In the electroweak--scale resonant leptogenesis scenario
\cite{Pilaftsis:2005rv} the Hubble scale at the time of 
leptogenesis corresponds to (roughly speaking) 
$\sim 10$ km at 0.1 MeV. At the latter temperature, which 
corresponds to the beginning of nucleosynthesis, the neutron 
diffusion length $\sim 200$ km. In other words, the baryon 
perturbations are not smoothed out on scales which correspond
to $\mathcal{X}\lesssim 0.1$ at the time of leptogenesis. 
Given the typical size of the integration constants 
$C_{1,2}\sim 10^{-5} - 10^{-4}$ the amplitude of the 
perturbations not washed out by baryon diffusion 
is too small to affect nucleosynthesis in an observable way. 

After the end of nucleosynthesis and before the beginning of   
recombination there are no neutral baryons and the diffusion
length substantially decreases. Thus the baryon isocurvature 
perturbations which have not been smoothed out survive till 
recombination -- the time of CMB formation. At 
this stage small scale perturbations blueof the 
photon gas undergo a mechanism called 
\emph{diffusion damping} or \emph{Silk damping} 
\cite{Silk:1967kq,Silk:1983}. The photons diffuse from regions 
with higher density to lower density regions and thereby 
effectively smooth out baryon inhomogeneities 
due to their large mean free path. This damps the amplitudes of 
acoustic oscillations by a factor
\begin{equation}
D \equiv \exp {\left( \int_{t_A} ^{t_0} \Gamma \text{d}x^0 \right)} \equiv \exp{ \left[ - \left( \frac{M_c}{M} \right) ^{2/3} \right] }, 
\end{equation}
where $\Gamma$ is the damping rate, $t^0$ the present time and 
$M_c$ is called the critical mass that can only be obtained 
approximately from the integral over the damping rate. For a 
radiation-dominated universe during the phase of acoustic 
oscillations, one obtains \cite{Boerner:1993}
\begin{equation}
M_c \approx 8\times 10^{13} (\Omega_0 h_0 ^2)^{-1/2} M_\odot 
\end{equation}
with density parameter $\Omega_0$ and rescaled Hubble parameter 
$h_0$ evaluated at the present time and $M_\odot$ denoting the 
solar mass. Sufficient damping is then obtained for $(M_c/M)$ of 
the order of 10 or larger. Translated into length scales, this 
means that all scales 
\begin{equation}
 \lambda_D \lesssim 3\times 10^{18} \text{m}
\end{equation}
are affected by Silk damping at $T \sim 10$ eV, which roughly 
corresponds to the temperature at recombination, and only 
perturbations on scales larger than that can survive.
For the right--handed neutrino mass $M_1\sim 10^9$ GeV, scales 
of the order of  $\cal R$ at the time of leptogenesis correspond 
to scales of the order of 10 m at recombination. This is obviously 
smaller than the damping scale $\lambda_D$  by so many orders of 
magnitude that no observable imprint on the CMB can be produced. 
Turning the argumentation around, we can calculate the mass scale 
$M_1$ that would be necessary to obtain perturbations on large enough 
scales to prevent them from being smoothed out by Silk damping, 
taking into account that $\mathcal{R}$ scales as $T^{-2}$ in a 
radiation--dominated universe. The result is very disappointing: 
one would need a mass scale of $M_1 \sim 100$ eV which is far 
below the scale where the sphalerons are in thermal equilibrium and 
can convert a lepton asymmetry into a baryon asymmetry. Therefore we 
conclude that in the strong-washout regime, the baryon inhomogeneities 
coming from the spatial fluctuation of the efficiency of leptogenesis 
can neither affect BBN nor the CMB in an observable way.

Given the estimate \eqref{retaweak2}, the conclusions drawn in the case of 
the strong washout regime also apply to the case of the weak 
washout regime.

\section{\label{conclusions}Summary and conclusions}

In this short paper we have investigated the influence of 
the primordial perturbations of the energy density and 
metric on the efficiency of thermal leptogenesis. 

To perform the analysis we have derived an integral 
form of the Boltzmann equation for a gauge--invariant 
combination $\Upsilon$ of particle number density, the 
macroscopic fluid velocity and space--time metric. In the
absence of the collision terms $\Upsilon$ is  conserved 
to linear approximation and replaces the particle number 
density per comoving volume in the inhomogeneous 
universe.

Using the integral form of the Boltzmann equation 
approximate analytical solutions for the lepton 
asymmetry in the weak and strong washout regimes 
have been found. 

In the strong washout regime the generated baryon isocurvature 
perturbations are correlated with the perturbations 
of the energy density and metric. 
In the weak washout regime the solution deviates 
from the background one only in  the presence of 
the heavy neutrino velocity perturbation $r_u$. 
The typical scale of the perturbations at the time of 
leptogenesis is set by the Hubble scale $\cal R$ in both
regimes. The amplitude is inversely proportional to the scale.

The generated perturbations are smoothed by baryon diffusion 
at the later stages of the universe's evolution. The scale
of the generated perturbations is comparable to the neutron diffusion 
length at the beginning of BBN only if the right--handed 
neutrino is as light as in the electroweak--scale resonant 
leptogenesis scenario. However even in this extreme case 
the generated perturbations are smoothed out by the \textit{Silk 
damping} at the recombination. For this reason the influence of 
the baryon inhomogeneities coming from the spatial fluctuation 
of the efficiency of leptogenesis on BBN and CMB is unobservably
small.

\begin{acknowledgments}
The authors are grateful to Prof.  E.  Paschos for drawing  their 
attention to the problem of leptogenesis in an inhomogeneous 
universe and numerous discussions, to Prof. D. Schwarz for his 
kind help and to Dr. T. Underwood for careful reading of the 
manuscript.
\end{acknowledgments}

\end{document}